\documentclass[12pt,preprint]{aastex}

\shorttitle{Ices toward Cepheus A East}
\shortauthors{Sonnentrucker et al.}

\begin{document}

%% LaTeX will automatically break titles if they run longer than
%% one line. However, you may use \\ to force a line break if
%% you desire.

\title{Fully Sampled Maps of Ices and Silicates in Front of Cepheus A East with {\it Spitzer}\altaffilmark{1}}

%% Use \author, \affil, and the \and command to format
%% author and affiliation information.
%% Note that \email has replaced the old \authoremail command
%% from AASTeX v4.0. You can use \email to mark an email address
%% anywhere in the paper, not just in the front matter.
%% As in the title, use \\ to force line breaks.

\author{P. Sonnentrucker\altaffilmark{2}, D. A. Neufeld\altaffilmark{2}, P. A. Gerakines\altaffilmark{3}, E. A. Bergin\altaffilmark{4}, G. J. Melnick\altaffilmark{5}, W. J. Forrest\altaffilmark{6}, J. L. Pipher\altaffilmark{6} and D. C. B. Whittet\altaffilmark{7}}

%% Notice that each of these authors has alternate affiliations, which
%% are identified by the \altaffilmark after each name.  Specify alternate
%% affiliation information with \altaffiltext, with one command per each
%% affiliation.

\altaffiltext{1}{Based on observations with the {\it Spitzer Space Telescope}.}
\altaffiltext{2}{Department of Physics and Astronomy, Johns Hopkins University, 3400 North Charles Street, Baltimore, MD 21218.}
\altaffiltext{3}{Department of Physics, University of Alabama at Birmingham, 310 Campbell Hall, 1300 University Blvd, Birmingham, AL 35294.}
\altaffiltext{4}{Department of Astronomy, University of Michigan, 825 Dennison Building, 
500 Church Street, Ann Arbor, MI 48109.}
\altaffiltext{5}{Harvard-Smithsonian Center for Astrophysics, 60 Garden Street, Cambridge, MA 02138.}
\altaffiltext{6}{Department of Physics \& Astronomy, University of Rochester, Rochester, NY 14627.}
\altaffiltext{7}{Department of Phyiscs, Applied Physics \& Astronomy, Rensselaer Polytechnic Institute, Troy, NY 12180.}

%% Mark off your abstract in the ``abstract'' environment. In the manuscript
%% style, abstract will output a Received/Accepted line after the
%% title and affiliation information. No date will appear since the author
%% does not have this information. The dates will be filled in by the
%% editorial office after submission.

\begin{abstract}

We report the first fully sampled maps of the distribution of interstellar CO$_2$ ices, H$_2$O ices and total hydrogen nuclei, as inferred from the 9.7 $\mu$m silicate feature, toward the star-forming region Cepheus A East with the IRS instrument onboard the {\it Spitzer Space Telescope}. We find that the column density distributions for these solid state features all peak at, and are distributed around, the location of HW2, the protostar believed to power one of the outflows observed in this star-forming region. A correlation between the column density distributions of CO$_2$ and water ice with that of total hydrogen indicates that the solid state features we mapped mostly arise from the same molecular clumps along the probed sight lines. We therefore derive average CO$_2$ ice and water ice abundances with respect to the total hydrogen column density of $X$(CO$_2$)$_{ice}$$\sim$1.9$\times$10$^{-5}$ and $X$(H$_2$O)$_{ice}$$\sim$7.5$\times$10$^{-5}$. Within errors, the abundances for both ices are relatively constant over the mapped region exhibiting both ice absorptions. The fraction of CO$_2$ ice with respect to H$_2$O ice is also relatively constant at a value of 22\% over that mapped region. A clear triple-peaked structure is seen in the CO$_2$ ice profiles. Fits to those profiles using current laboratory ice analogs suggest the presence of both a low-temperature polar ice mixture and a high-temperature methanol-rich ice mixture along the probed sightlines. Our results further indicate that thermal processing of these ices occurred throughout the sampled region. 
\end{abstract}

%% Keywords should appear after the \end{abstract} command. The uncommented
%% example has been keyed in ApJ style. See the instructions to authors
%% for the journal to which you are submitting your paper to determine
%% what keyword punctuation is appropriate.

\keywords{ISM: Molecules --- ISM: Clouds, molecular processes --- Star forming region: individual (Cepheus A East)}

%% From the front matter, we move on to the body of the paper.
%% In the first two sections, notice the use of the natbib \citep
%% and \citet commands to identify citations.  The citations are
%% tied to the reference list via symbolic KEYs. The KEY corresponds
%% to the KEY in the \bibitem in the reference list below. We have
%% chosen the first three characters of the first author's name plus
%% the last two numeral of the year of publication as our KEY for
%% each reference.

%% Authors who wish to have the most important objects in their paper
%% linked in the electronic edition to a data center may do so by tagging
%% their objects with \objectname{} or \object{}.  Each macro takes the
%% object name as its required argument. The optional, square-bracket 
%% argument should be used in cases where the data center identification
%% differs from what is to be printed in the paper.  The text appearing 
%% in curly braces is what will appear in print in the published paper. 
%% If the object name is recognized by the data centers, it will be linked
%% in the electronic edition to the object data available at the data centers  
%%
%% Note that for sources with brackets in their names, e.g. [WEG2004] 14h-090,
%% the brackets must be escaped with backslashes when used in the first
%% square-bracket argument, for instance, \object[\[WEG2004\] 14h-090]{90}).
%%  Otherwise, LaTeX will issue an error. 

\section{Introduction}

Understanding how the structure and composition of ice mantles covering interstellar grains change with the local physical and chemical conditions is crucial to constraining the chemical evolution of protostellar envelopes, protoplanetary disks, and comets. Infrared observations toward low-to-high mass star-forming regions (e.g., Nummelin et al. 2001; Gibb et al. 2004) and quiescent molecular clouds (Whittet et al. 1998) using the {\it Infrared Space Observatory} have shown that water (H$_2$O), carbon monoxide (CO) and carbon dioxide (CO$_2$) are common constituents of the ice mantles covering interstellar grains in dense molecular environments. While H$_2$O and CO are believed to form from grain surface reactions and gas-phase desorption, respectively (e.g., d'Hendecourt et al. 1985), the production of solid CO$_2$ remains uncertain. Mechanisms such as UV photolysis and/or grain surface chemistry are invoked. The routine detection of solid CO$_2$, with a fraction relative water ice between 9 and 37 \% toward a number of star-forming regions (e.g., Gerakines et al. 1999; Nummelin et al. 2001), combined with the fact that CO$_2$ is easily produced via UV irradiation in the laboratory (Ehrenfreund et al. 1996), added weight to the suggestion that CO$_2$ ices are mainly produced via UV photolysis in the interstellar medium. However, the detection of CO$_2$ ice toward a number of quiescent clouds devoid of any embedded source of radiation, with abundances relative to water ice similar to those measured toward star-forming regions, demonstrated that production mechanisms such as grain surface reactions and gas-grain interaction need be considered too (Whittet et al. 1998; Whittet et al 2007).

A number of laboratory experiments studied the evolution of ice mixtures when subjected to thermal and/or UV processing (e.g., Ehrenfreund et al. 1996; 1999). Comparisons of these laboratory ice analogs with {\it ISO} and more recent {\it Spitzer} observations of interstellar ices allowed previous workers to better constrain the physical and chemical conditions in the clouds toward which these ices were detected (e.g., Ehrenfreund et al. 1998; 1999). In particular, the profile of the $\nu_2$ vibrational bending mode of solid CO$_2$ (15.2 $\mu$m) was found to be very sensitive to the composition of the ice mantle the molecules are embedded in and to the thermal history of the region under study.  Substructures that are characteristic of ice crystallization and segregation of the ice mantle constituents upon thermal processing have been routinely used to constrain the physical and chemical environments around young stellar objects (e.g., Boogert et al. 2000; Gerakines et al. 1999; Boonman et al. 2003; Gibb et al. 2004; Bergin et al. 2005). These previous studies were all based on pointed observations of individual discrete sources, and thus require observations of multiple sources to provide any spatial information. With the {\it Spitzer} spectral mapping capability, one can now investigate the distribution of interstellar ices within a given dense cloud or star-forming region with unprecedented spatial sampling ($\sim$3$''$). This provides a new opportunity to link the spectral properties of the ices with local physical conditions, and thus to throw new light on their origin and evolution. In this paper, we report on the spatial distribution and the evolution of the interstellar ices present toward the star-forming region Cepheus A East using this new capability offered by {\it Spitzer}.

Cepheus A is a well-known site of star formation located at a distance of about 650 pc. The region contains a series of deeply embedded far-infrared and radio-continuum sources, one of which dominates the luminosity of the entire region (HW2 with 2.5$\times$10$^{4}$ $L_{\odot}$; Hughes \& Wouterloot 1984; Lenzen et al. 1984; Evans et al. 1981). Ground- (e.g., Hartigan et al. 1996; Goetz et al. 1998) and space-based observations (e.g., Wright et al. 1996; van den Ancker et al. 2000) revealed the existence of a multipolar outflow exhibiting complex structures of shock-excited atomic and molecular gas components, indicating the presence of both dissociative (J-type) and non-dissociative (C-type) shocks resulting from successive episodes of activity (Narayanan \& Walker 1996). The protostellar object HW2 was found to be the dominant powering source of the extremely high-velocity (EHV) outflow oriented {\it northeast} ({\it NE}), while the source (or sources) of the high-velocity jet (HV; oriented {\it southeast}) is still under debate (e.g., Goetz et al. 1998; Hiriart et al. 2004; Codella et al. 2003; Mart\'{\i}n-Pintado et al. 2005). 

Previous observations using {\it ISO} exhibited a number of absorption bands due to the presence of ice mantles containing CO, CO$_2$ and H$_2$O  and due to silicate grains over the {\it NE} side of this star forming region (van den Ancker et al. 2000). The CO and CO$_2$ ice abundances -- with respect to water ice and averaged over the Short-Wavelength Spectrometer (SWS) aperture -- were found to be within the range observed toward other sight lines (e.g., Whittet et al. 1996). Using new high-resolution {\it Spitzer} data, we recently reported the first detection of gas-phase CO$_2$ emission. This emission extends over a 35$''\times$25$''$ region associated with the EHV outflow (Sonnentrucker et al. 2006). We determined that the gaseous CO$_2$ molecules mostly result from sputtering of ice mantles due to interactions between the EHV outflow and the ambient molecular medium. We derived a CO$_2$ gas-to-ice ratio of at most 3\% over the region showing CO$_2$ gas emission. 

Our data also exhibit strong absorption from the 15 $\mu$m CO$_2$ ice bending mode, from the 6 $\mu$m water ice band and from the 9.7 $\mu$m silicate feature at the position of HW2 and at numerous spatial positions preferentially located away from the EHV outflow region. In this paper, we discuss the distribution of the solid state features observed toward Cepheus A East with the {\it Spitzer} Infrared Spectrograph (IRS). While H$_2$ $S$(0) to $S$(7), [NeII], [NeIII], [S I], [SIII] and [FeII] emissions as well as emission from C$_2$H$_2$ (Sonnentrucker et al., submitted) are also detected in our data, we focus here on the study of the absorptions from CO$_2$ ices, H$_2$O ices and silicate grains -- the latter used to estimate the total column density of hydrogen nuclei $N$(H$_{\rm{tot}}$) -- toward Cepheus A East. Section 2 summarizes our observations and Section 3 describes our data analysis. Section 4 compares the spatial distribution of the CO$_2$ ice,  H$_2$O ice and silicate grains with that of the quiescent molecular clouds present in this region. Section 5 discusses the ice abundances as well as the variations in ice mantle composition derived from fits to representative CO$_2$ ice profiles using current laboratory ice analogs.

%% In a manner similar to \objectname authors can provide links to dataset
%% hosted at participating data centers via the \dataset{} command.  The
%% second curly bracket argument is printed in the text while the first
%% parentheses argument serves as the valid data set identifier.  Large
%% lists of data set are best provided in a table (see Table 3 for an example).
%% Valid data set identifiers should be obtained from the data center that
%% is currently hosting the data.
%%
%% Note that AASTeX interprets everything between the curly braces in the 
%% macro as regular text, so any special characters, e.g. "#" or "_," must be 
%% preceded by a backslash. Otherwise, you will get a LaTeX error when you 
%% compile your manuscript.  Special characters do not 
%% need to be escaped in the optional, square-bracket argument.

\section{Observations and data reduction}

Spectral maps of two overlapping 1$'$ $\times$ 1$'$-square fields were obtained toward Cepheus A East with the IRS instrument onboard the {\it Spitzer Space Telescope (SST)} as part of Guaranteed Time Observer (GTO) program 113. The Short-Low (SL, both orders), Short-High (SH) and Long-High (LH) modules allowed for wavelength coverage from 5.2 to 35 $\mu$m. Since the flux density at 25 $\mu$m exceeds the limit of 50 Jy causing severe detector saturation, most of the LH data were not used. Continuous spatial coverage in the fields was obtained by stepping the slit perpendicular and parallel to its long axis in steps of one-half its width and 4/5 its length, respectively. 

The data were processed at the Spitzer Science Center (SSC) using version 12 of the IRS pipeline. We used the Spectroscopy Modeling Analysis and Reduction Tool (SMART) software (Higdon et al. 2004) as well as  a set of locally-developed IDL routines (Neufeld et al. 2006) to extract wavelength-calibrated spectra, to remove bad pixels in the high resolution modules, to calibrate the fluxes for extended sources, and to generate spectral line maps from the extracted spectra. 

\section{Data analysis}

Absorption caused by solid state features such as water ice (6.02 $\mu$m), amorphous silicates (9.7 $\mu$m) and CO$_2$ ice (15.2 $\mu$m) are clearly seen in most spectra. To compare the distribution of these ices with that of the gas-phase species known to result from interactions between the Cepheus A East outflows and the ambient molecular gas (e.g., Goetz et al. 1998; van den Ancker et al. 2000; Sonnentrucker et al. 2006), we derived the CO$_2$ ice, H$_2$O ice and total hydrogen column densities as described below. Note that the absorption feature at 6.85 $\mu$m has an origin which remains uncertain (e.g., Keane et al 2001; Gibb \& Whittet 2002; Gibb et al. 2004); while certainly present in our spectra, this feature is not discussed here.  

\subsection{CO$_2$ and H$_2$O column densities}

We derive the column densities ($N_m$) of solid CO$_2$ (15.20 $\mu$m) and H$_2$O (6.02 $\mu$m) by dividing the integrated optical depth profiles, in wavenumbers, by their respective molecular band strengths ($A_m$); $N_m=1/c\,\int \tau(\nu)d\nu/A_m$. We used third-degree polynomials to fit the local continuum around the 15 $\mu$m CO$_2$ ice bending mode and to derive the corresponding optical depth profiles.  We adopted a band strength of $A_{CO_2}=$ 1.1$\times$10$^{-17}$ cm molecule$^{-1}$ (Gerakines et al. 1995). Figure 1 displays an example of continuum fitting ({\it upper panel}) and optical depth profile generation ({\it lower panel}) for the CO$_2$ ice absorption observed in a summed spectrum at the position of the protostellar source HW2. This summed spectrum was obtained by adding the individual spectra contained in a 6$''\times$8$''$ region centered on HW2. In the {\it upper panel}, the thick smooth curve overimposed on the summed spectrum represents the best fit to the local continuum at this spatial position. In the {\it lower panel}, the thick horizontal segment shows the wavelength range over which the CO$_2$ optical depth profile was integrated to derive the corresponding CO$_2$ column density.   

Because of the complex structure present in the continuum around the 6 $\mu$m water ice absorption feature, we constrained the local continuum fit using two narrow wavelength ranges that appeared to be the least affected by intervening species (from 5.2 to 5.7 $\mu$m and from 7.4 to 7.6 $\mu$m). To produce the water ice optical depth profiles, we fitted these continuum windows using a linear combination of a first-order polynomial and a Gaussian function with fixed width and fixed line center to account for emission arising from the H$_2$ $S$(7) line at 5.52 $\mu$m. We adopted a H$_2$O band strength of $A_{H_2O}=$ 1.2$\times$10$^{-17}$ cm molecule$^{-1}$ (Gerakines et al. 1995) to derive the water ice column density in the mapped region. 

To estimate the uncertainties in column density introduced by our fits to the continuum local to the CO$_2$ and H$_2$O ice features, we first calculated the standard deviation ($\sigma$) around our best fit continuum in each continuum window. We then shifted the best-fit continuum by $\pm$1 $\sigma$ and generated the corresponding optical depth and new estimate of the ice column densities. The difference between the best-fit column density and the columns obtained by shifting the best-fit continuum model constitutes our $\pm$1 $\sigma$ error in $N$(CO$_2$)$_{ice}$ and $N$(H$_2$O)$_{ice}$. The dashed lines in the {\it upper panel} of Fig.~1 show the shifted model continuum we used to estimate the $\pm$1 $\sigma$ uncertainty in the CO$_2$ ice column density for this particular summed spectrum. We used that same technique to derive the 1 $\sigma$ uncertainty in the H$_2$O ice column densities.

\subsection{Hydrogen column density}

Because direct measurements of the total hydrogen column density are not possible toward the region we mapped, we estimated the total column density of hydrogen nuclei by fitting the observed 9.7 $\mu$m silicate absorption feature (SL1 data alone) with a linear combination of (i) a first-order polynomial, (ii) the renormalized synthetic Galactic extinction curve per unit hydrogen column calculated for $R_{\rm V}=A_{\rm V}/E_{\rm B-V}=$ 5.5 (Weingartner \& Draine 2001; Li \& Draine 2001; Draine 2003a\& b) and, (iii) three Gaussian functions with fixed widths and centroids to account for emission from the H$_2$ $S$(4), H$_2$ $S$(2) and [\ion{Ne}{2}] lines ($\lambda$8.02, 12.24 and 12.81 $\mu$m, respectively). Because numerous silicate features show signs of saturation in their cores toward Cepheus A East, we restricted our fits to the wings of the observed silicate profiles (from 7.7 to 8.95 $\mu$m and 12 to 14 $\mu$m). Note also that our fitting model assumes that most of the dust resides in a cold foreground component with no significant contribution of background emission from hot dust. If there were underlying silicate emissions produced by hot dust in addition to the absorption due to cold dust, then one should regard our hydrogen column density estimates as lower limits to the true columns along the probed sight lines containing hot dust (Gillett et al. 1975).    Figure~2 displays examples of best fits (thick red curves) to the observed 9.7 $\mu$m silicate profile (black curves) at two spatial positions: the region around the protostellar source HW2 believed to power the EHV outflow ({\it upper panel}) and the region around two radio-continuum sources (HW5 \& HW6) where interaction between the EHV outflow and the ambient material is believed to take place ({\it lower panel}; see also Goetz et al. 1998; Sonnentrucker et al. 2006). 

We also produced an additional set of fits to the observed silicate profiles using the synthetic Galactic extinction curve per unit hydrogen column calculated for $R_{\rm V}=$3.1 in order to assess the impact that dust properties have on our hydrogen column density estimates. We find that the hydrogen columns derived assuming dust properties consistent with $R_{\rm V}=$3.1 are systematically greater by $\sim$18\% than those derived assuming $R_{\rm V}=$5.5. Thus, while the dust properties adopted to fit the observed silicate features do affect the overall hydrogen column density scale we derived, these assumptions do not affect the local spatial variations that we see in the $N$(H$_{\rm{tot}}$) distribution shown in Fig.~3, assuming that the silicate profiles scale with $N$(H$_{\rm{tot}}$). Since a number of studies suggested that the average Galactic extinction curve calculated for $R_{\rm V}=$ 5.5 is most appropriate to describe the extinction properties of dense molecular clouds (e.g., Sharpless 1952; Vrba \& Rydgren 1984; Whittet et al. 2001), we adopted the total hydrogen column densities obtained with the dust extinction properties corresponding to $R_{\rm V}=$ 5.5 for the remainder of our study. With the latter dust grain model, the extinction per hydrogen column has a value of 4.957$\times$10$^{-23}$ cm$^2$/H at 9.7 $\mu$m (Draine 2003a \& b).

To estimate the uncertainties in the total hydrogen column density introduced by the adopted silicate profile fitting function, as before we calculated the standard deviation ($\sigma$) around our best-fit model to the observed silicate feature in a wavelength range between 13 and 13.95 $\mu$m in each case. We then increased the best-fit hydrogen column, $N$(H$_{\rm{tot}}$), by $\pm$ 1 $\sigma$, $\pm$ 2 $\sigma$ and $\pm$ 3 $\sigma$ and compared these new fits to the observed silicate profiles. Our comparison indicates that a $\pm$ 1 $\sigma$ error of at least 5\% in the derived $N$(H$_{\rm{tot}}$) -- as shown by the blue dash-dotted lines in Fig.~2 -- often adequately reflects the uncertainties related to our silicate absorption model fits. We adopted these $\pm$ 1 $\sigma$ errors on the hydrogen column densities in the subsequent sections when discussing the ice abundance distribution in Cepheus A East (see Figs. 5 \& 6). Note that Chiar et al. (2007, submitted) recently reported the strength of the silicate feature to be systematically weaker relative to near-IR extinction in molecular clouds compared to diffuse clouds. Hence, an additional systematic source of error in our estimate of the total hydrogen column density could also arise from our adopted Galactic dense cloud extinction curve model.

%% In this section, we use  the \subsection command to set off
%% a subsection.  \footnote is used to insert a footnote to the text.

%% Observe the use of the LaTeX \label
%% command after the \subsection to give a symbolic KEY to the
%% subsection for cross-referencing in a \ref command.
%% You can use LaTeX's \ref and \label commands to keep track of
%% cross-references to sections, equations, tables, and figures.
%% That way, if you change the order of any elements, LaTeX will
%% automatically renumber them.

%% This section also includes several of the displayed math environments
%% mentioned in the Author Guide.

\section{Results}

Figure~3 shows the column density distributions we obtained toward the portion of the Cepheus A East map common to both the SL and SH modules for CO$_2$ ice ({\it upper left}), water ice ({\it upper right}) and total hydrogen -- as inferred from the silicate absorption feature -- ({\it lower left}). The gray contours display the spatial distribution of NH$_3$(1,1) emission concentrated into 3 major clumps of cold quiescent molecular gas, CepA-1, -2 and -3 (Torrelles et al. 1993). The coordinates are offsets in Right Ascension and Declination with respect to HW2 (J2000.0: $\alpha=$22$^{h}$56$^{m}$17$^s$.9 and $\delta= +$62$^{\circ}$01$'$49$''$; Hughes \& Wouterloot 1984), the powering source of the EHV outflow (e.g., Goetz et al. 1998), in units of arcseconds. The crosses indicate the positions of some of the known radio-continuum sources in this region (Hughes \& Wouterloot 1984). The rectangles ({\it upper right}) show examples of spatial regions over which individual IRS spectra were summed to generate the regions labeled a,b, P1 and P2 (see Sections 5.2 and 5.3 below).
 
In general, the CO$_2$ ice, H$_2$O ice and total hydrogen column density distributions peak at, and closely around, the position of the still deeply embedded protostellar source HW2 where substantial extinction ($A_{\rm{V}}>$75 mag; Lenzen et al. 1984) allows for a significant ice build-up associated with the NH$_3$(1,1) clump labeled CepA-1 to occur (Torrelles et al. 1993). There seems to be no clear association between the ice distributions and the two other NH$_3$(1,1) clumps CepA-2 and CepA-3 where outflow interactions have been previously reported (e.g., Goetz et al. 1998; Sonnentrucker et al. 2006). Additionally, local differences in the ice distributions seem to exist, the most striking being the very patchy and somewhat more extended structures observed for the CO$_2$ ices compared to those observed for the H$_2$O ices and the silicates around the CepA-1 clump as well as the weak water ice absorption compared to CO$_2$ ice over the region labeled P2.

The {\it lower right panel} of Fig.~3 displays the background emission distribution, in logarithmic scale, derived from a linear fit to the continuum local to the CO$_2$ ice feature and centered at 14.95 $\mu$m. This map (together with the 7.5 $\mu$m continuum map not displayed here) shows that continuum dust emission is present throughout the observed region. The overall spatial variations exhibited by the silicate and the two ice maps are, therefore, not due to a lack of dust continuum emission against which to absorb background light. Previous observations established the presence of an IR reflection nebulosity located a few arcseconds {\it north} of HW2, which scatters the radiation emanating from the HW2 protostellar region (e.g., Lenzen et al. 1984; Goetz et al. 1998). The extent and peak position of the dust emission we measure are consistent with those of this reflection nebulosity, indicating that the dust continuum emission shown here originates from that same region.

Figure~4 displays examples of summed spectra in the 5.2-17 $\mu$m range at five positions indicated in the maps of Fig.~3. We generated these data by adding all individual spectra included in a $\sim$ 6$''\times$ 8$''$ region centered on the marked positions in order to increase the lines S/N ratio and minimize effects due to the difference in spatial resolution between the low- and the high-resolution data. Low-resolution summed spectra from 5.2 to 11 $\mu$m (SL, 2 orders) were combined with corresponding summed high-resolution spectra from 11 $\mu$m on (SH), in each case. These five spectra were chosen to be representative of the local spatial variations observed in both the gas-phase and the solid state content. In particular, note the complete absence of gas-phase emission features but the presence of strong solid state absorption features at the HW2 position, in contrast with the rich gas-phase spectrum and the weak ice features at the $NE$ position. A comparison of positions P1 and P2 further illustrates potential local spatial variations in the CO$_2$ ice and H$_2$O ice contents, as already suggested by the maps of Fig.~3. Finally, variations in the 15.2 $\mu$m CO$_2$ ice profile are also seen and are indicative of spatial changes in the ice mantle composition and/or in the thermal history of this star forming region (see Section 5.4 and Gerakines et al. 1999; Gibb et al. 2004).

\section{Discussion}

While absorption from solid H$_2$O and solid CO$_2$ were previously measured toward this region using {\it ISO} (e.g., van den Ancker et al. 2000), the unparalleled {\it Spitzer} spatial resolution and its much greater sensitivity allowed us both to map and compare the distribution of these ices for the very first time over a much larger spatial extent than was previously possible. Because the H$_2$O ice, the total hydrogen and the CO$_2$ ice column densities are obtained from different modules (SL and SH, respectively), we produced a set of contiguous summed spectra over the region mapped in both SL and SH in order to minimize the effects that different spatial resolution might have on the single-pixel extracted spectra and to increase the line S/N ratios. A total of 28 such summed spectra were produced, five of which are shown in Fig.~4. It is the column densities we derived from these contiguous summed spectra that we use to discuss the ice content toward Cepheus A East in greater detail below. 

\subsection{CO$_2$ ice abundance}

Previous observations shed light on the complex molecular structures existing toward Cepheus A East. In particular, molecular clumps of hot gas ($T\sim$730-1000K; van den Ancker et al. 2000; our data), warm gas ($T\sim$50-250 K; Sonnentrucker et al. 2006) and much colder gas ($T\sim$20-50K; Torrelles et al. 1993; G\'omez et al. 1999) are clearly seen in some of our spectra by means of emission from H$_2$ $S$(3) to $S$(7), of gas-phase CO$_2$ and H$_2$ $S$(2), and of NH$_3$(1,1) emission or absorption from the ices, respectively. While constraining the spatial distribution of these various molecular clumps is often difficult because the quantities we derive are averaged over the probed sight lines, our data however indicate that the hot gas traced by the H$_2$ ($J>$3) emissions and the warm gas traced by the gas-phase CO$_2$ and H$_2$ $S$(2) emissions clearly arise in front of the bulk of the colder molecular clumps traced by the ices and the silicate absorption features. Otherwise emission from H$_2$ $S$(3) at 9.66 $\mu$m would not be detected against the saturated silicate absorption features seen at positions P2 or HW5/6, for instance (see Fig.~4). Consequently, the CO$_2$ ice abundance cannot be derived using column density measurements obtained from the observed H$_2$ lines toward this region. 

Alternatively, we can derive the ice abundances from our estimates of the total hydrogen column density obtained by fitting the silicate absorption profiles, if all species mostly coexist spatially along the probed sight lines. To determine whether the CO$_2$ ice and the silicate grain absorptions arise predominantly from the same molecular clumps, we compared the CO$_2$ and the total hydrogen column density distributions over the region mapped both in SL and SH. Figure~5 displays $N$(CO$_2$)$_{ice}$ $vs$ $N$(H$_{\rm{tot}}$) for the 28 contiguous summed spectra covering the SL and SH maps. Within errors, the CO$_2$ ice column density increases with increasing total hydrogen column, as expected if both species arise predominantly from the same cold molecular clumps. Since some scatter does exist, we calculated the Pearson correlation coefficient to determine how significant this apparent correlation might be. With our sample size, a correlation coefficient of at least $r=$ 0.48 would indicate that the probability for this correlation to be due to chance is less 1\%. We find a correlation coefficient of $r=$ 0.83. Hence, our data show that the CO$_2$ ice and the hydrogen column densities do increase together, thereby confirming that the two species predominantly arise from the same cold molecular clumps. 

Recent observational results obtained toward intraclump molecular clouds (i.e. clouds devoid of embedded sources of radiation) confirmed that a minimum visual extinction $A_{\rm{V}}=$ 4.3 $\pm$ 1.0 mag (corresponding to $N$(H$_{\rm{tot}}$) $=$ 8.2 $\times$ 10$^{21}$ cm$^{-2}$; see Bohlin et al. 1978; Whittet et al. 2001) seems required for substantial CO$_2$ ice to build up onto dust grains (e.g., Bergin et al. 2005; Whittet et al. 2007; see also Chiar et al. 1995). To determine whether this ''threshold effect" pertains toward the Cepheus A East region, we searched for the linear combination that best fitted our data points. The best-fit combination to our measurements (black line in Fig.~5) has the form $N({\rm CO_2})_{ice} = (1.9\pm0.4) \times 10^{-5}  [N({\rm H}_{\rm{tot}}) - N_0]$ with $N_0=$ (2.6 $\pm$ 2.2) $\times$ 10$^{22}$ cm$^{-2}$ (1 $\sigma$ error bars). Considering our large uncertainties, the extinction threshold obtained from our best-fit line, while apparently higher, is not inconsistent with the extinction threshold determined previously. 

In light of the overall good $N$(CO$_2$)$_{ice}$--$N$(H$_{\rm{tot}}$) correlation exhibited by our data, we derived the CO$_2$ abundance -- $X$(CO$_2$)$_{ice}$ -- by dividing $N$(CO$_2$)$_{ice}$ by its corresponding $N$(H$_{\rm{tot}}$) toward our contiguous summed spectra. Because CO$_2$ ice has been shown to build-up on grains once the extinction threshold is reached, the slope of the linear relationship displayed in Fig.~5 essentially measures the average CO$_2$ ice abundance for the region we sampled with {\it Spitzer} toward Cepheus A East. We therefore find that the CO$_2$ ice distribution shows a mean abundance of $X$(CO$_2$)$_{ice}=$ (1.9 $\pm$ 0.4)$\times$ 10$^{-5}$ which is consistent with abundances either measured previously (e.g., van Dishoeck et al. 1996; Boogert et al. 2004) or predicted by chemical models allowing for efficient surface diffusion to occur (e.g., Charnley 1997; Ruffle \& Herbst 2001).

\subsection{H$_2$O ice abundance}

{\it ISO} observations of the 3.05 and 6.02 $\mu$m water ice bands toward a sample of sight lines comprising quiescent molecular clouds as well as young stellar objects (YSOs) of intermediate and high masses showed that the column density derived from the 6 $\mu$m water ice band was systematically overestimated by up to a factor 3 with respect to the column derived from the 3 $\mu$m band toward YSOs (e.g., Keane et al. 2001; Gibb \& Whittet 2002). The excess absorption coinciding with the 6 $\mu$m  water ice band was attributed to an intervening species the origin of which remains unclear todate. The selective appearance of this additional absorption feature toward star-forming regions, however, led to the conjecture that these unidentified species might result from some energetic processing (Gibb \& Whittet 2002). 

Because of the {\it Spitzer} wavelength cut-off, the water ice column densities we derived from our data (see Fig.~3, {\it upper right}) rely solely on the 6 $\mu$m band measurements and, hence, might potentially be overestimated if such blending were to occur in our spectra. Van den Ancker et al. (2000) observed part of the Cepheus A East with the {\it ISO}/SWS 14$''\times$27$''$ aperture centered on the infrared source IRS6a (Hughes \& Wouterloot 1984). IRS6a is located at ($\Delta\alpha\cos\delta=$ $+$5.3$''$; $\Delta\delta=$ $+$11$''$), i.e. 4  arc-seconds North of the radio-continuum source HW4 in our maps. In other words, the region over which the average {\it ISO} spectrum was obtained is included in our larger {\it Spitzer} map and covers most of the CepA-2 and -3 NH$_3$(1,1) clouds as well as the EHV outflow region. A comparison of our water ice measurements based on the 6 $\mu$m band alone with those obtained by van den Ancker et al. (2000) from both water ice features shows very good consistency over this common region within errors. Most importantly, van den Ancker et al. measured no difference between the 3 and 6 $\mu$m water ice column densities over this region, implying that no significant blending of the 6 $\mu$m ice feature occurs over the CepA-2, -3 and EHV outflow regions. 

In order to determine whether significant blending might occur over the CepA-1 NH$_3$(1,1) cloud where the water ice column is the highest in our data, we compared the water ice and total hydrogen column density distributions derived from our summed spectra. Figure 6 displays $N$(H$_2$O)$_{ice}$ $vs$ $N$(H$_{\rm{tot}}$) for the 28 contiguous summed spatial positions. Our data suggest that the water ice and the total hydrogen column increase in concert in the mapped region. For our sample, the Pearson correlation coefficient for water ice is $r=$0.76, much higher than the minimum correlation coefficient required to ensure a chance probability lower than 1\%. Our data, hence, indicate that the water ice and the hydrogen column densities do increase together thereby implying that the two species predominantly arise from the same cold molecular clumps (as found for CO$_2$ ices). 

As for CO$_2$ ice, we calculated the linear combination that best fitted our data points for water ice. The best-fit combination to our measurements (black line in Fig.~6) has the form $N({\rm H_2O})_{ice} = (7.5 \pm 1.7) \times 10^{-5}  [N({\rm H}_{\rm{tot}}) - N_0]$ with $N_0=$ (2.3 $\pm$ 2.5) $\times$ 10$^{22}$ cm$^{-2}$ (1 $\sigma$ error bars). Departures from the best linear fit to our data are noticeable for three spatial regions marked a, b and P1 (see Fig.~6). Inspection of the summed spectra indicates that these positions correspond to the peaks in H$_2$O ice column seen in Fig.~3 ({\it upper right}), at the location of the CepA-1 NH$_3$(1,1) clump. To our knowledge, no data containing the 3 $\mu$m water ice band are available over this particular region. We are, thus, unable to determine whether the local enhancements we observe in the water ice columns are real or due to blending with unidentified species. As in the case of the CO$_2$ ices, the best-fit line to the H$_2$O ice-total hydrogen relationship shows an intercept consistent (within large errors) with the water ice extinction threshold of 3.2 $\pm$ 0.1 mag and corresponding to $N$(H$_{\rm{tot}}$) $=$ 6.1 $\times$ 10$^{21}$ cm$^{-2}$, determined by Whittet et al. (2007).

In light of the good correlation between $N$(H$_2$O)$_{ice}$ and $N$(H$_{\rm{tot}}$), we derived the water ice abundance -- $X$(H$_2$O)$_{ice}$-- by dividing $N$(H$_2$O)$_{ice}$ by its corresponding total hydrogen column density in all cases. The slope of the best-fit line  displayed in Figure~6 indicates that the H$_2$O ice distribution shows an average abundance of $X$(H$_2$O)$_{ice}=$ (7.5 $\pm$ 1.7) $\times$ 10$^{-5}$, a value are  consistent with those observed elsewhere (e.g., Nummelin et al. 2001; Boogert et al. 2004) over the region mapped both in SL and SH. Within errors, the H$_2$O ice abundance excesses observed for the regions labeled a, b and P1 again correspond to $N$(H$_2$O)$_{ice}\ge$ 9 $\times$ 10$^{18}$ cm$^{-2}$, i.e. the high-end of the water ice column density distribution where blending with unidentified features is most likely to have occurred. 
 
\subsection{Ice evolution toward Cepheus A East}

The formation, destruction and evolution of ices are believed to be governed by relatively few processes intimately linked to the local physical conditions: 1) energetic processes involving UV photolysis and cosmic-ray bombardment; 2) grain surface reactions; 3) thermal processing of the ice mantles by embedded protostars; and 4) sputtering in shocks. Because CO$_2$ ices are readily formed via UV photolysis in the laboratory, this formation mechanism was favored to account for the routine detection of CO$_2$ ices toward star-forming regions. However, the detection of relatively high fractions of CO$_2$ ice relative to H$_2$O ice toward quiescent clouds and the intra-cloud medium indicated that CO$_2$ ices can be formed in environments devoid of embedded sources of radiation with a typical fraction with respect to water ice of 17\% (e.g., Whittet et al. 1998; Bergin et al. 2005; Whittet et al. 2007). Thus, CO$_2$ ice formation mechanisms, other then UV photolysis, need be considered (e.g., Charnley et al. 1997; Ruffle \& Herbst 2001). 

Figure 7 compares the column density distribution of CO$_2$ ices $versus$ that of water ice for the 28 contiguous summed spectra common to the SL and SH maps of Cepheus A East. The thick black line represents the typical ISM fraction of CO$_2$ ice with respect to water ice of 17\% (Gerakines et al. 1999). The thin black line shows the best fit to our measurements excluding regions a, b and P1 where potential significant blending of the 6 $\mu$m water ice feature might occur. A slope of 22 $\pm$ 3\% fit our data points best. We find a Pearson correlation coefficient of 0.77. For our sample size, this value is much greater than the minimum coefficient of 0.48 required for the correlation to be significant at a 1\% level. Our data, therefore, indicate that the CO$_2$ ice and the water ice build up in concert onto dust grains under very similar physical conditions and that they essentially coexist in the same ice mixture. Our results strengthen the case that these two species formed together as discussed by Bergin et al. (2005), with a CO$_2$ ice fraction with respect to water ice overall similar to those toward the typical quiescent molecular clouds studied so far (e.g., Whittet et al. 2007). Therefore, grain surface reactions rather than UV photolysis seem to be the most likely CO$_2$ ice production mechanism toward most parts of the star-forming region mapped in this study.

However, note that local CO$_2$ ice enhancements with respect to water ice are clearly observed in our data, mostly around P2 and around the region centered at ($+$2;$-$16) (see Fig.~3), suggesting the existence of additional sources of CO$_2$ ice. Variations in the CO$_2$ ice fraction compared to the typical ISM value were reported previously toward other sources (e.g., Whittet et al. 1998; Nummelin et al. 2001; Boogert et al. 2004) and mechanisms such as UV or/and thermal processing were invoked. In the present case, the local CO$_2$ ice excesses we observe coincide mostly with the region where a high-velocity (HV) jet interacts with the CepA-3 molecular clump.  Goetz et al. (1998) reported the presence of compact \ion{H}{2} regions along this (HV) jet providing local sources of UV radiation. One might, thus, conjecture that the CO$_2$ excesses with respect to H$_2$O ice we measure at the latter positions result from photochemistry involving CO and O$_2$ (e.g., Sandford \& Allamandola 1990; Whittet et al. 1996). Since we do not possess information concerning the CO ice distribution in the region mapped with {\it Spitzer} we are, however, unable to further test this possibility at present.

\subsection{Thermal history of Cepheus A East}

Laboratory experiments have shown that, upon heating, ice mixtures with various concentrations of H$_2$O, CO$_2$ and CH$_3$OH undergo crystallization leading to an effective segregation of the ice constituents in the mixtures (e.g., Ehrenfreund et al. 1997; 1999). Observationally, such a phenomenon is detectable through the appearance of multiple substructures in the profile of the CO$_2$ ice bands. Such multiple peaked structures can not be produced by UV irradiation in the laboratory (Ehrenfreund et al. 1999). The profile of the CO$_2$ $\nu_2$ bending mode (15.2 $\mu$m) has been shown to be the most sensitive to the ice mantle composition and to the mantle thermal processing by nearby protostars. In particular, the presence of a sharp double-peaked structure at $\sim$15.2 $\mu$m along with a broader feature around 15.4 $\mu$m in the CO$_2$ profile, has been recognized as a signature of thermal processing of ice mixtures containing roughly equal amounts of H$_2$O, CO$_2$ and CH$_3$OH. The long-wavelength shoulder is characteristic of segregation of CO$_2$ and CH$_3$OH in the ice mantle upon heating (e.g., Ehrenfreund et al. 1998). Comparisons of the observed 15.2 $\mu$m band profiles with laboratory interstellar ice analogs have subsequently led to a better understanding of the composition of interstellar ice mantles and have allowed us to constrain the local molecular environment these ice mantles are subjected to (e.g., Gerakines et al. 1999; Nummelin et al. 2001; Boogert et al. 2004; Bergin et al. 2005).

All CO$_2$ bands we mapped with sufficient S/N ratio exhibit a sharp double-peak substructure along with a broad long-wavelength shoulder of variable intensity, as shown in Fig.~4. To compare our observations with the currently available laboratory interstellar ice analogs (Ehrenfreund et al. 1997, 1999), we derived the CO$_2$ optical depth profiles by fitting a third-degree polynomial to the local continuum around the 15 $\mu$m band and we adopted the fitting method described in Gerakines et al. (1999) to constrain the ice mantle composition toward Cepheus A East. Because the CO$_2$ ice bands expand over two SH orders which overlap around 15.2 $\mu$m, we only retained those spectra with the highest S/N ratio and no apparent order-edge mismatches to ensure that the most accurate fits to both the band wings and the sharp substructures are obtained. Examples of such optical depth spectra are shown in Figure 8 (black curves) along with their respective best laboratory ice analog fits (red curves).

In all cases, our best fits require two ice components at two different laboratory temperatures, a broad polar component (H$_2$O:CO$_2$:CO 100:20:3) at 20 K (blue curve in Fig.~8) and a triple-peaked methanol-rich mixture (H$_2$O:CH$_3$OH:CO$_2$ 1:1:1) at 119 K (green curve in Fig.~8)\footnote{Note that the laboratory temperature of 119K required to fit our data effectively corresponds to a much lower temperature in space since, in the ISM, the relevant parameter is the time scale for ice segregation rather than the temperature. For instance, if the time scale for ice segregation were of the order of 10$^{5}$ years toward Cep A East, then the 119K laboratory temperature we obtain from our fits effectively corresponds to an ISM segregation temperature of $\sim$76K (e.g. Boogert et al. 2000).}. From these best fits we estimate that the fraction of low-temperature ice present in the mantles ranges from 25 \% at the position of HW2 (Fig.~8, upper panel) to 40\% over the CepA-2 region (Fig.~8, lower panel). To evaluate the robustness of our ice mixture solution and because the optical depth spectra are very sensitive to the continuum normalization, we produced a second dataset using a spline function to fit the local continuum. We fitted this second set of optical depth spectra with the same ice mixtures and laboratory temperatures but allowing the cold-ice fraction to vary. The range in cold-ice fraction which led to these new best fits is reported in Figure 8 for each spectrum. Within errors, our data indicate that the concentration of low-temperature ice is of at least 30\%  over the 4 probed positions which were chosen to be representative of the ice composition over the mapped Cepheus region. Local variations in the low-temperature ice fraction seem to exist with a slight increase of this concentration when moving toward the CepA-2 clump and away from the {\it NE} outflow activity (2 lower panels of Fig.~8). The need for an increasingly large fraction of a high-temperature ice mixture containing methanol in the fits indicates that the ice mantles experienced significant thermal processing over the history of the region, leading to segregation of the CO$_2$ and methanol ices in the mantles.

\section{Conclusion}

We used new {\it Spitzer} data obtained with the IRS instrument to produce fully sampled maps of the distribution of CO$_2$ ices (15.2 $\mu$m), H$_2$O ices (6.02 $\mu$m) and total hydrogen nuclei, as inferred from the 9.7 $\mu$m silicate feature, toward the star-forming region Cepheus A East. We find that all solid state features peak at, and are distributed closely around, the spatial position of the deeply embedded protostar HW2 and coinciding with the NH$_3$(1,1) molecular clump CepA-1. Otherwise, the distributions show column densities lower by about a factor 2 than those measured over the CepA-1 region.
  
The correlations we observe between the CO$_2$ ice, the H$_2$O ice column density distributions and that of total hydrogen imply that the solid state features predominantly arise from the same molecular clouds along the probed sight lines. We hence derived the CO$_2$ and water ice abundances with respect to the measured total hydrogen column density at each summed spatial position in the map. We find an average CO$_2$ ice abundance of (1.9 $\pm$ 0.4) $\times$ 10$^{-5}$ and an average H$_2$O ice abundance of (7.5 $\pm$ 1.7) $\times$ 10$^{-5}$ over the probed region. 

A comparison of the CO$_2$ and water ice column density distributions indicates that both ices build up onto dust grains, in concert, over the probed region. The fairly good correlation between the column densities of the two ices also indicates that blending of the water ice band by unidentified features is not significant in our data.  Overall, we find that $N$(CO$_2$)$_{ice}=$ (0.22 $\pm$ 0.03) $\times$ $N$(H$_2$O)$_{ice}$, a fraction similar to that typically found in the intra-could medium and toward quiescent molecular clouds (17 \%), suggesting that grain surface chemistry is the most likely production mechanism of CO$_2$ ices toward this region. 

Best fits to the CO$_2$ ice bending mode using the current laboratory interstellar ice analogs database indicate that the ice mantles in Cepheus A East are composed of 2 ice mixtures, a low-temperature ($T_{lab}$=20K) polar ice mixture and a high-temperature ($T_{lab}$=119K) methanol-rich mixture. We find that while about 30\% of the probed ice mantles is at low temperature, variations of the cold-ice fraction also seem to exist over the probed region.
The presence of the high-temperature methanol-rich ice mixture is indicative of significant thermal processing of a fraction of these ice mantles over Cepheus A East history.

\acknowledgments

This work, which was supported in part by JPL contract 960803 to the Spitzer IRS Instrument Team and by RSA agreement 1263841, is based on observations made with the {\it Spitzer Space Telescope}, which is operated by the Jet Propulsion Laboratory, California Institute of Technology under a NASA contract. We are grateful to J.D. Green and K-H Kim for helping with the SMART software and standard IRS reductions. D.A.N. and P.S. acknowledge funding from LTSA program NAG 5-13114 to the Johns Hopkins University. We are grateful to J.M. Torrelles for providing us with the NH$_3$ map. We thank the referee for useful comments.

\clearpage

\begin{figure}[hf]
\epsscale{.70}
\plotone{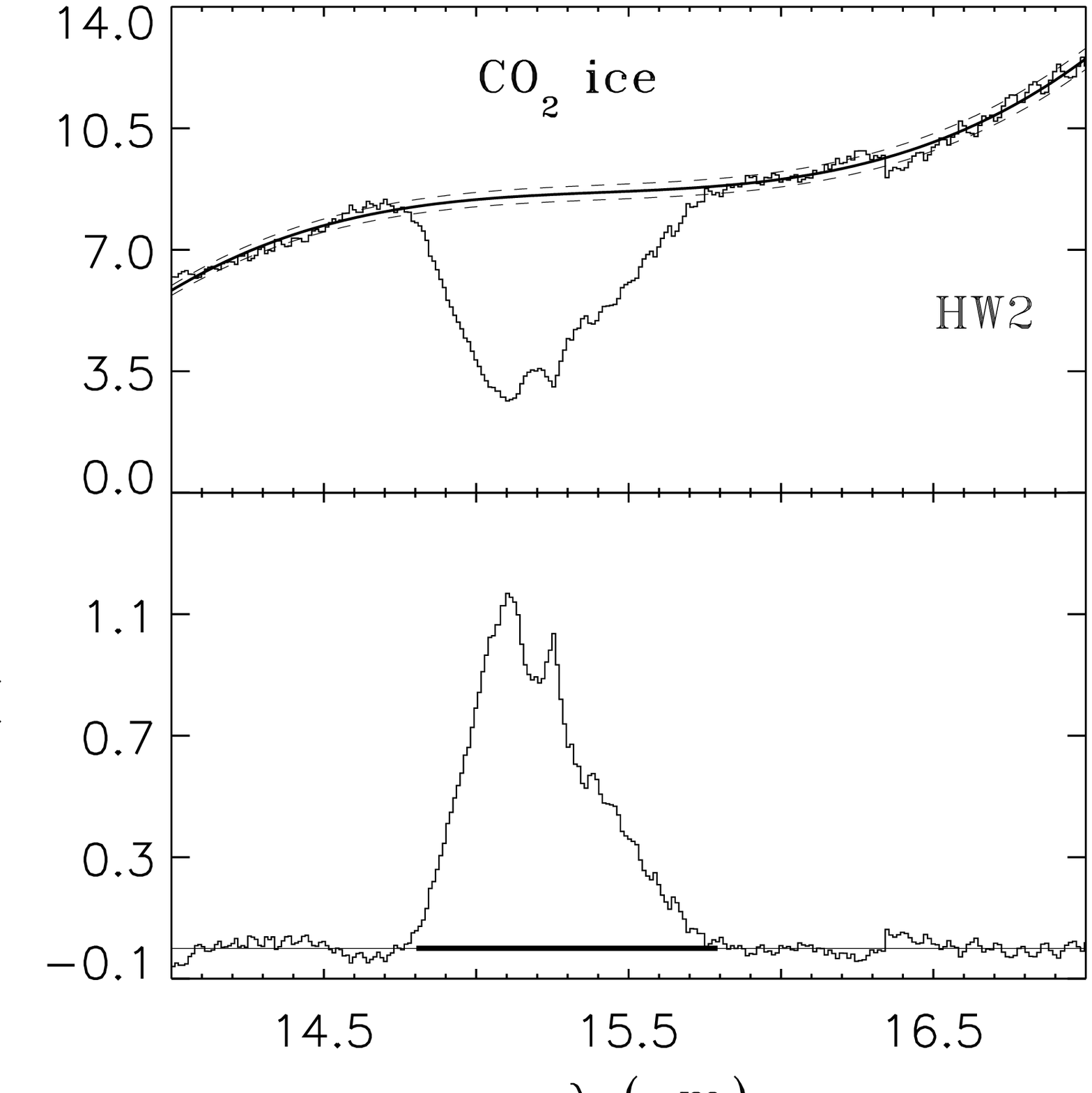}
\caption{{\it Upper panel:} CO$_2$ ice summed spectrum at the position of HW2 (J2000.0: $\alpha=$22$^{h}$56$^{m}$17$^s$.9 and $\delta= +$62$^{\circ}$01$'$49$''$; Hughes \& Wouterloot, 1984). The thick smooth curve shows the best-fit continuum used to derive the optical depth profile with $\tau(\lambda)=$$-\log I(\lambda)_{\rm{obs}}/I(\lambda)_{\rm{cont}}$. The thin dashed lines show the continuum best fit shifted by $\pm$ 1 $\sigma$ used to estimate the CO$_2$ ice column density uncertainties. {\it Lower panel:} Resulting optical depth profile for the best fit continuum. The thick black segment delineates the wavelength range over which the optical depth integration was performed in all cases. \label{fig1}}
\end{figure}

\begin{figure}[hf]
\epsscale{.70}
\plotone{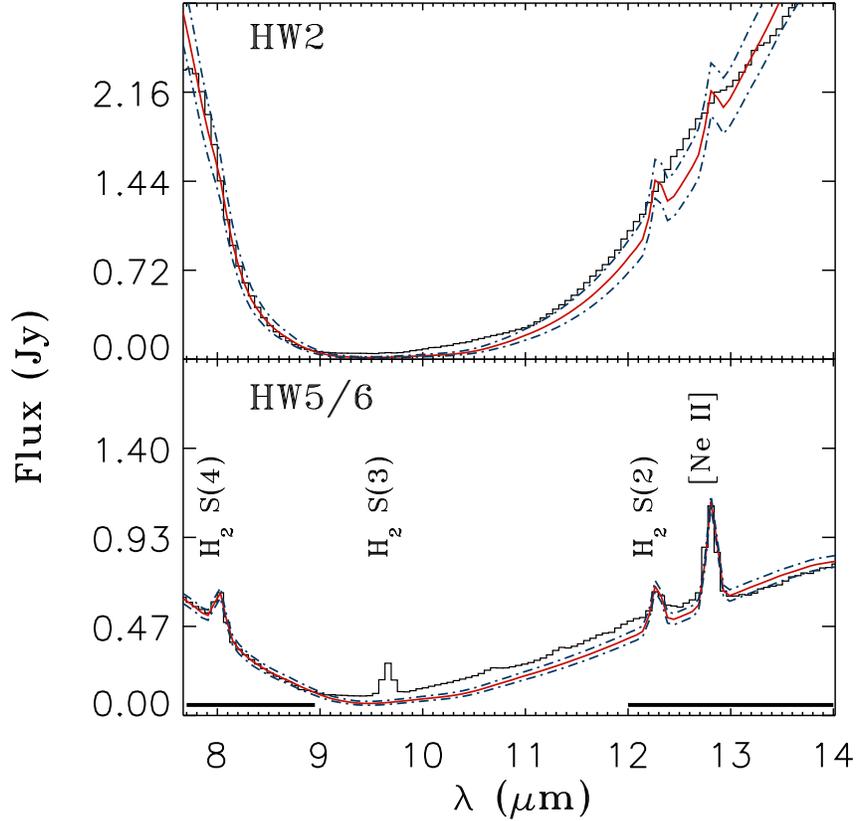}
\caption{Summed spectrum at the position of the radio-continuum sources HW2 ({\it upper panel}) and HW5/6 ({\it lower panel}, see also Fig.~3). The thick red curves show the best-fit to the silicate feature ($\lambda$9.7 $\mu$m) for these two particular regions using the synthetic Galactic extinction curve per hydrogen column calculated for $R_{\rm{V}}=$ 5.5 (Draine 2003a \& b). The thick horizontal lines ({\it lower panel}) indicate the wavelength ranges used to constrain the fits using SL data alone. We used this technique to derive the total hydrogen column density -- $N$(H$_{\rm{tot}}$) -- at all spatial positions mapped in SL. The blue dash-dotted lines indicate the $\pm$ 1 $\sigma$ uncertainty in our fit to the observed silicate profile. \label{fig2}}
\end{figure}

\begin{figure}[hf]
\epsscale{1.0}
\plottwo{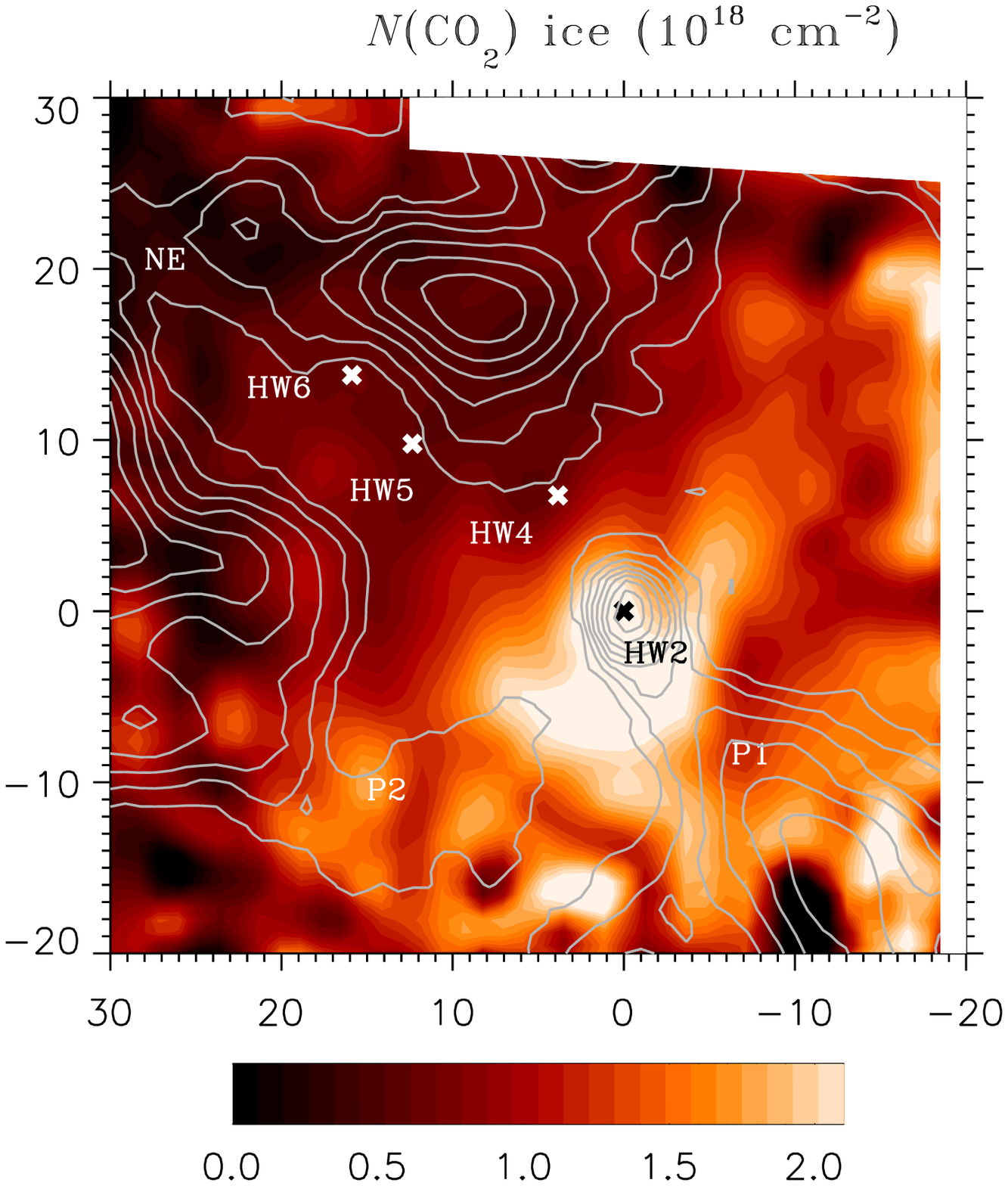}{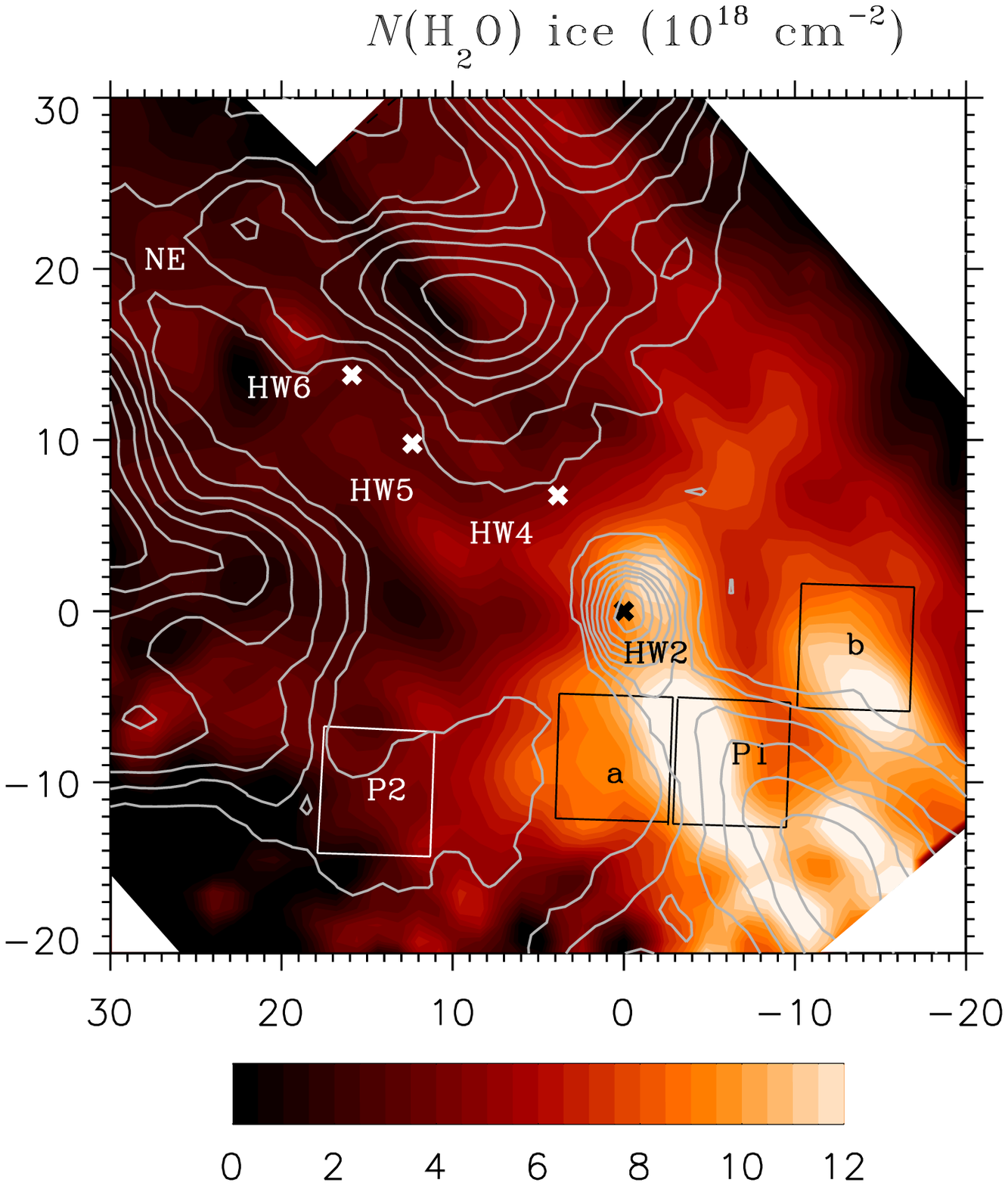}\\
\plottwo{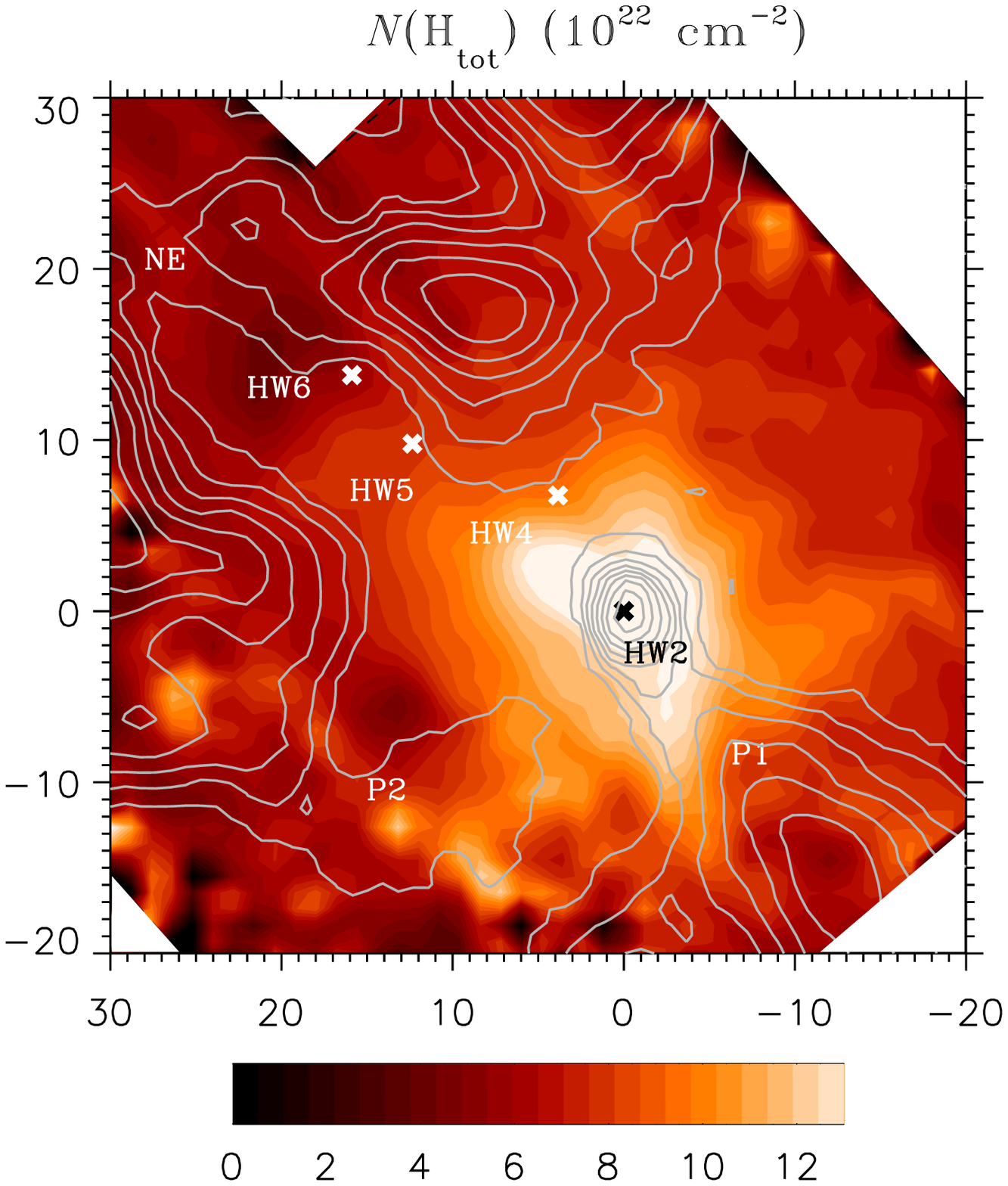}{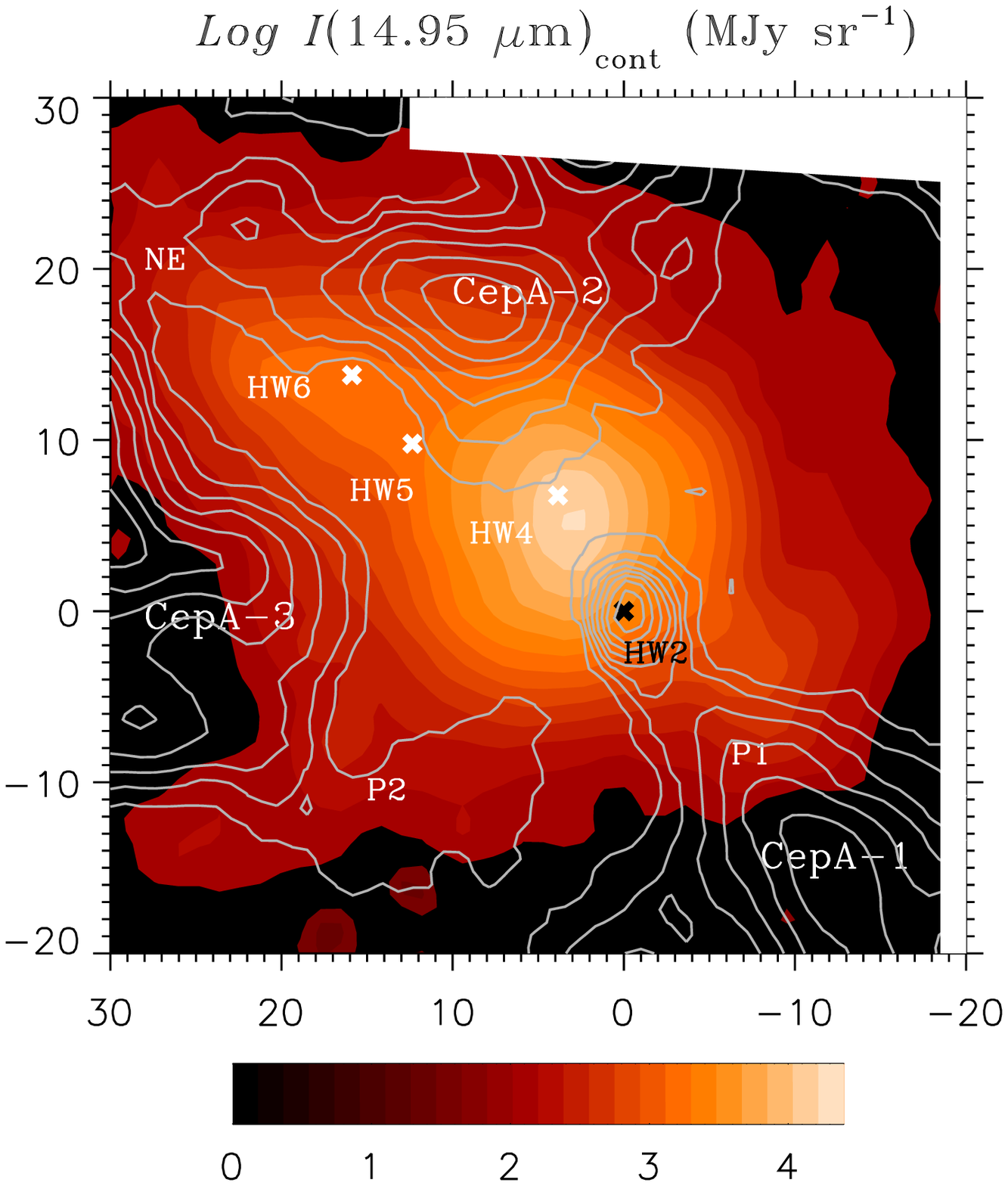}
\caption{{\it Upper left:} Column density distribution of solid CO$_2$ ($\lambda$15.2 $\mu$m) absorption. {\it Lower left:} column density distribution of silicates ($\lambda$9.7 $\mu$m) tracer of $N$(H$_{\rm{tot}}$). {\it Upper right:} Column density distribution of H$_2$O ice ($\lambda$6.02 $\mu$m). {\it Lower right:} Continuum emission distribution at 14.95 $\mu$m. The crosses indicate the positions of some of the known radio-continuum sources in this region. The coordinates are offsets in arcseconds in Right Ascension and Declination with respect to HW2. The gray contours show the distribution of NH$_3$(1,1) into 3 clumps of cold quiescent molecular gas, CepA-1, -2 and -3 (Torrelles et al. 1993). The lowest contours are 10 and 25 in steps of 25 mJy km s$^{-1}$ beam$^{-1}$. The rectangles ({\it upper right}) show examples of spatial regions over which individual IRS spectra were combined to generate the summed IRS spectra labeled a, b, P1 and P2. These regions are further discussed in Section 5.2 and 5.3. The summed spectra at P1 and P2 are shown in Fig.~4. \label{fig3}}
\end{figure}

\begin{figure}[hf]
\epsscale{.70}
\plotone{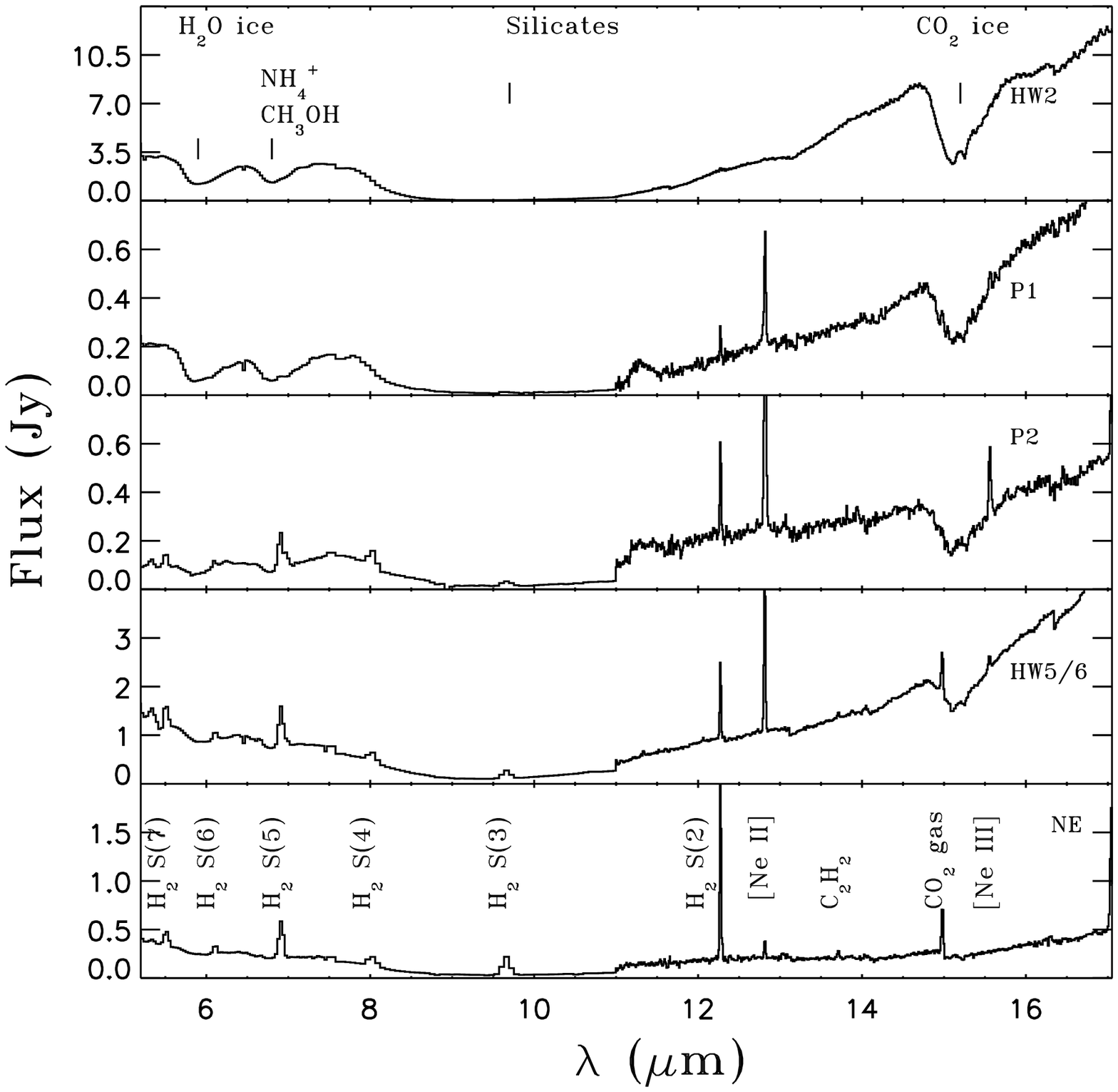}
\caption{Summed IRS spectra of CO$_2$ ice, H$_2$O ice and silicate absorption and gas-phase CO$_2$, H$_2$, C$_2$H$_2$, [Ne II] and [Ne III] emission toward Cepheus A East. Low-resolution data (from 5.2 to 11 $\mu$m) were combined with their corresponding high-resolution data (from 11 to 19 $\mu$m) in each panel. The individual spectra were summed over $\sim$ 6$''\times$ 8$''$ regions centered on the radio-continuum sources HW2, HW5/6, the {\it northeast} ($\Delta\alpha\cos\delta=$ $+$27$''$; $\Delta\delta=$ $+$22$''$) position, and the P1 ($\Delta\alpha\cos\delta=$ $-$6$''$; $\Delta\delta=$ $-$9$''$) and P2 ($\Delta\alpha\cos\delta=$ $+$15$''$; $\Delta\delta=$ $-$11$''$) positions, respectively (see Fig.~3). 28 such contiguous summed spectra were generated to cover the portion of the map common to both SL and SH and search for correlations among the solid state features detected toward these sightlines (see Figs.~5-7). \label{fig4}}
\end{figure}

\begin{figure}[hf]
\epsscale{0.70}
\plotone{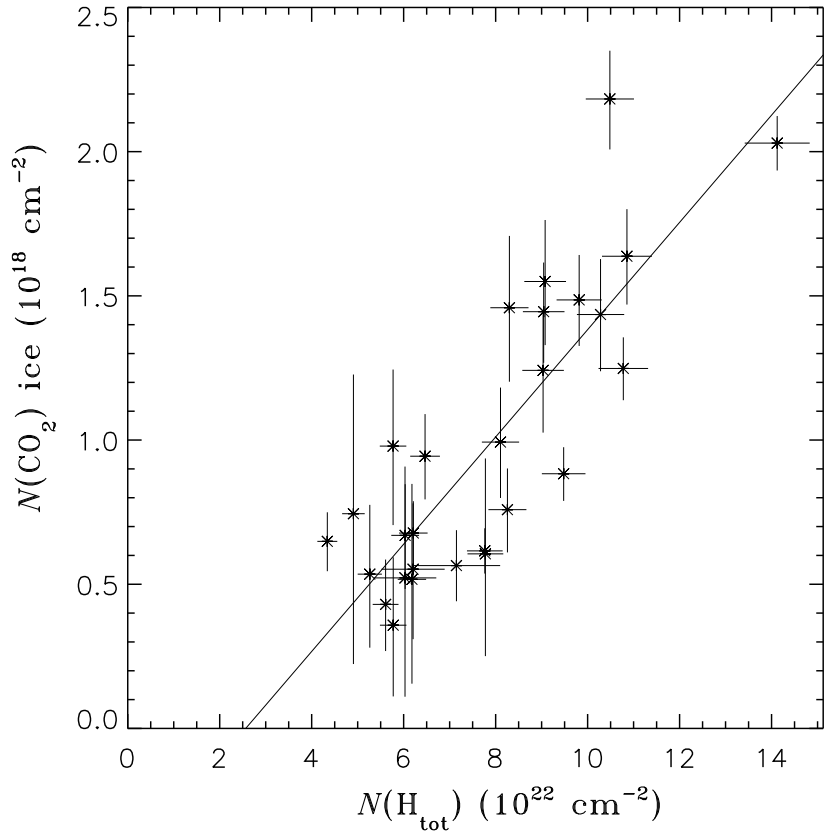}
%{sections_abun_co2_ice_Rv5.eps}\\
\caption{CO$_2$ ice column density {\it versus} total hydrogen column density as derived from fits to the 9.7 $\mu$m silicate features. The best-fit linear relation (solid line) has the form $N({\rm CO}_2)_{ice} = (1.9\pm 0.4)\times 10^{-5}[N({\rm H}_{\rm{tot}}) - (2.6 \pm 2.2) \times 10^{22}]$  (1 $\sigma$ error bars). Once a minimum total hydrogen column density is reached, the observed distribution shows a fairly constant average abundance $X$(CO$_2$)$_{ice}=$ (1.9 $\pm$ 0.4) $\times$ 10$^{-5}$ with respect to $N$(H$_{\rm{tot}}$) over the extent of the region common to SL and SH observations. \label{fig5}}
\end{figure}

\begin{figure}[hf]
\epsscale{.70}
\plotone{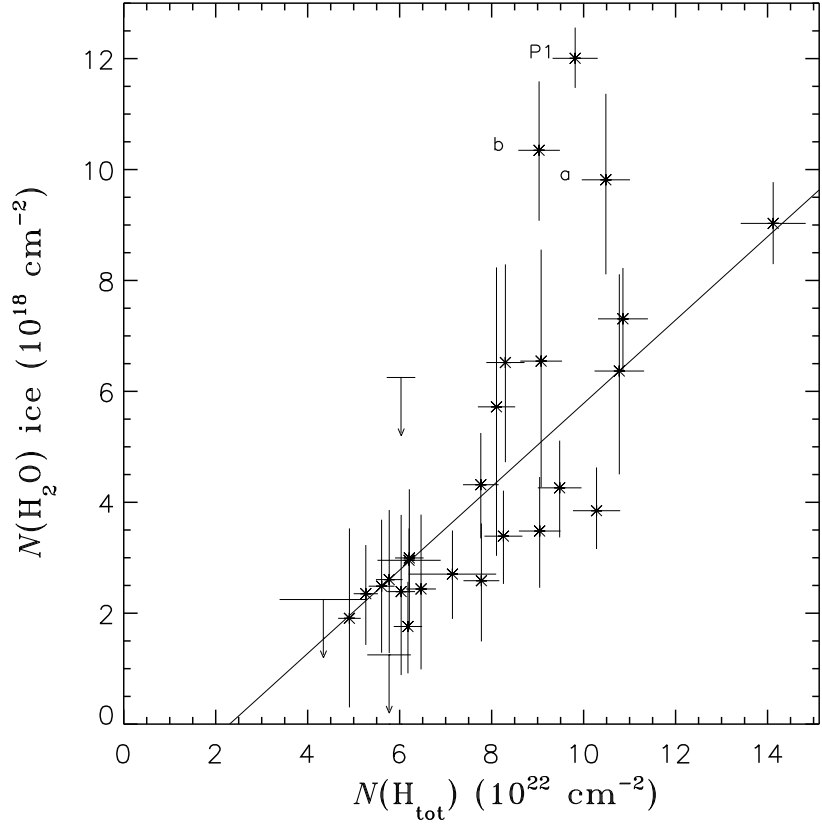}
%{sections_abun_h2o_ice_Rv5.eps}
\caption{H$_2$O ice column density $versus$ total hydrogen column density. The best linear fit (solid line) to our data has the form $N({\rm H_2O})_{ice}= (7.5 \pm 1.7) \times 10^{-5}[N({\rm H}_{\rm{tot}}) - (2.3 \pm 2.5) \times 10^{22}]$ (1 $\sigma$ error bars). As for water ice, once a minimum total hydrogen column density is reached the observed CO$_2$ distribution shows a fairly constant average abundance $X$(H$_2$O)$_{ice}=$ (7.5 $\pm$ 1.7) $\times$ 10$^{-5}$ with respect to $N$(H$_{\rm{tot}}$) over the spatial region common to SL and SH observations. Departures from the mean distribution can be seen for the regions labeled a, b and P1. Our current data do not allow us to determine whether these variations are real or due to blending of the water ice band with unidentified features.\label{fig6}}
\end{figure}

\begin{figure}[hf]
\epsscale{.70}
\plotone{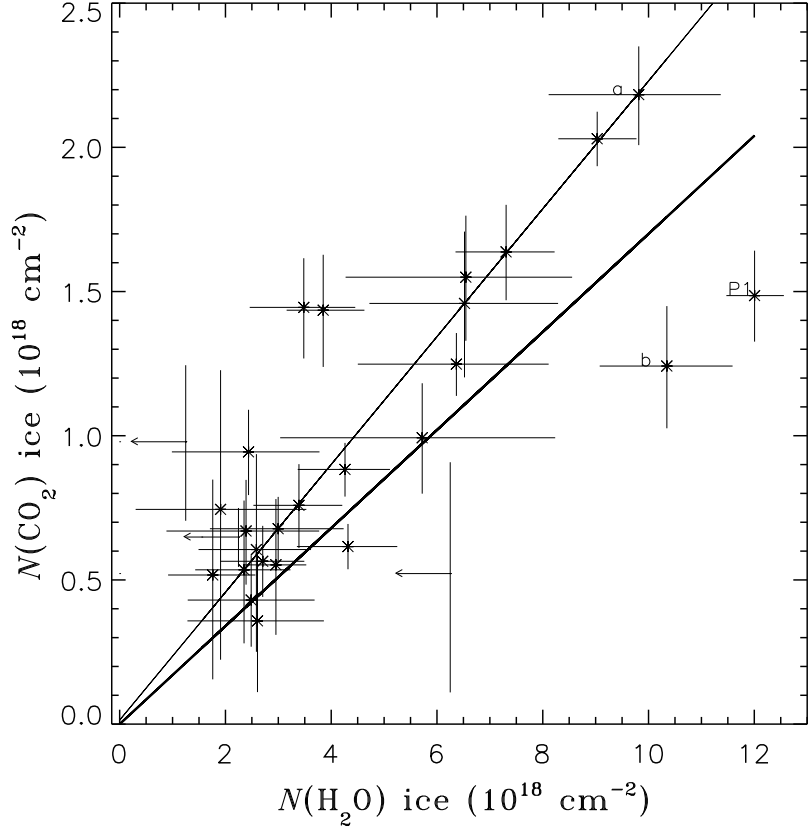}
\caption{Column density of CO$_2$ ice {\it versus} H$_2$O ice. The linear relation that best fits our measurements excluding regions a, b and P1 (see Section 5.3), is shown as a thin black line and suggests that \hbox{$N$(CO$_2$)$_{ice}=$ (0.22 $\pm$ 0.03) $\times$ $N$(H$_2$O)$_{ice}$}. The thick black line corresponds to a slope of 0.17, typical of cold quiescent molecular clouds (e.g., Gerakines et al. 1999; Whittet et al. 2007). Because the water ice columns might be locally overestimated due to potential blending with unidentified features (see Section 5.2), the average value of 0.22 derived from our data might only represent a lower limit to the fraction of CO$_2$ ice present at some positions such as regions b and P1.\label{fig7}}
\end{figure}

\begin{figure}[hf]
\epsscale{.70}
\plotone{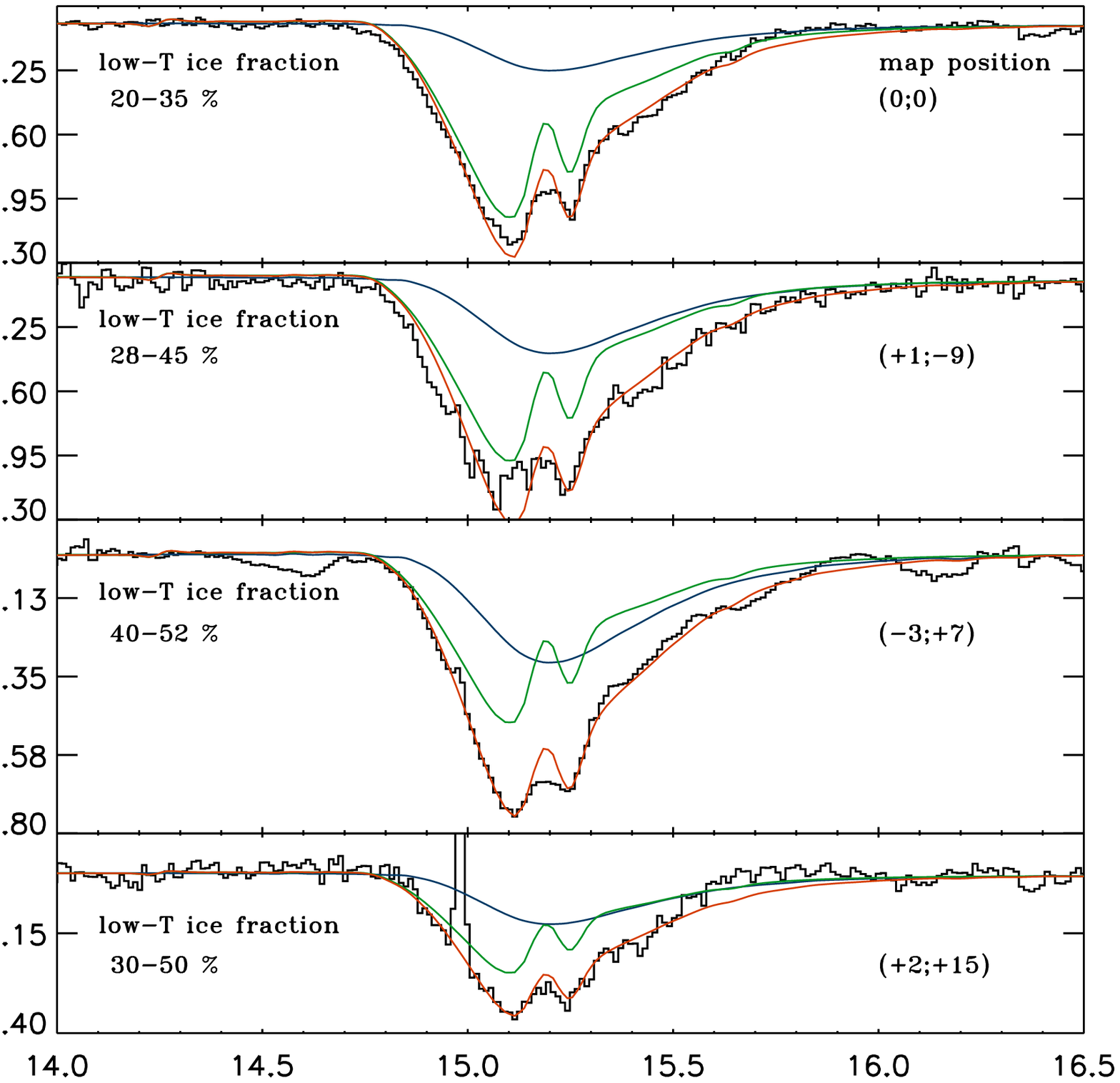}
\caption{Examples of CO$_2$ ice profiles and corresponding best-fits using current laboratory ice analogs. The black curves show the CO$_2$ optical depth spectra at four representative positions reported in the upper right of each panel (see Fig.~3). The red curves show the best fit to the data using a 2-component, 2-temperature ice mixture. The blue curves represent the $T_{lab}$=20 K polar, methanol-free ice mixture contribution to the overall profiles while the green curves exhibit the contribution of the high-temperature ($T_{lab}$=119K) methanol-rich ice mixture to the observed spectra. The relative concentration of low-temperature ice mixture is reported in the upper left of each panel. While the ice mantles at all probed positions seem to contain about 30\% of cold ices in their composition, spatial variations of this cold-ice fraction are also suggested by our data. \label{fig8}}
\end{figure}

\end{document}